# Dynamic Investment Portfolio Optimization under Constraints in the Financial Market with Regime Switching using Model Predictive Control

Vladimir Dombrovskii[a], Tatyana Obyedko[b]

Tomsk State University, Department of Economics and Department of Applied Mathematics and Cybernetics, Tomsk, Russia

[a] E-mail: dombrovs@ef.tsu.ru
[b] E-mail: tatyana.obedko@mail.ru

**Abstract**
In this work, we consider the optimal portfolio selection problem under hard constraints on trading volume amounts when the dynamics of the risky asset returns are governed by a discrete-time approximation of the Markov-modulated geometric Brownian motion. The states of Markov chain are interpreted as the states of an economy. The problem is stated as a dynamic tracking problem of a reference portfolio with desired return. We propose to use the model predictive control (MPC) methodology in order to obtain feedback trading strategies. Our approach is tested on a set of a real data from the radically different financial markets: the Russian Stock Exchange MICEX, the New York Stock Exchange and the Foreign Exchange Market (FOREX).

## 1. Introduction

The investment portfolio (IP) management is an area of both theoretical interest and practical importance. The foundation for modern portfolio selection theory is the single-period mean-variance approach suggested by Markowitz (1952) and the Merton's (1990) IP model in continuous time. At present, there exists a variety of models and approaches to the solution of the IP optimization problem.

In recent years considerable interest has been focused on regime-switching, or Markov-modulated, models to describe the behavior of the economic systems (Cogley and Sargent, 2009; Hansen, Mayer and Sargent, 2010; Lanne, Lütkepohl and Maciejowska, 2010). Models with Markov jumps are successfully used to describe the price dynamics of the risky asset for investment (Hamilton and Susmel, 1994; Elliott, Malcolm and Tsoi, 2003). This is due to the fact that these models can explain some important features of real financial markets.

Motivated by the great importance of these models from both theoretical and practical perspectives, optimization techniques to various portfolio selection problems under Markov regime-switching models have been intensively studied in the literature (see, for instance, Elliott and Van der Hoek (1997), Billio and Pelizzon (2000), Honda (2003), Ang and Bekaert (2002), Guidolin and Timmermann (2007), Gerasimov and Dombrovskii (2003), Bäuerle and Rieder (2004), Yin and Zhou (2004), Galperin et al. (2005), Costa and Araujo (2008), Sotomayor and Cadenillas (2009), Yiu et al. (2010), Dombrovskii and Obyedko (2011), Liu (2011), Wu (2013)).

In Honda (2003), Ang and Bekaert (2002), Guidolin and Timmermann (2007), Liu (2011) dynamic optimal consumption and portfolio choice problem, where expected returns of a risky asset follow a hidden Markov chain is studied. In Yin and Zhou (2004) a Markowitz's mean-variance portfolio selection with regime-switching problem is presented. The proposed target function looks only at the final variance of the portfolio. Costa and Araujo (2008) consider a multi-period generalized mean-variance model with Markov switching in the key market parameters. In this paper, all intermediate values of the mean and of the variance of the portfolio are involved in the objective. In Yiu et al. (2010) the authors investigate the optimal portfolio selection problem subject to a maximum value-at-risk constraint where the expected return and the volatility of the risky asset switch over time according to a continuous-time Markov chain. Sotomayor and Cadenillas (2009) studied an optimal investment problem with the bankruptcy constraint under a Markovian regime-switching model for the asset price dynamics. Wu (2013) investigates a non-self-financing mean-variance portfolio selection model in a Markov regime-switching jump-diffusion market with a stochastic cash-flow. In Gerasimov and Dombrovskii (2003), Galperin, Dombrovskii and Fedosov (2005) the portfolio selection problem in the financial market with regime switching is stated as a dynamic tracking problem of a reference portfolio with desired return under quadratic performance criterion.

The vast majority of the existing literature on dynamic portfolio selection just based on the using dynamic programming approach for determining the solution. However, that approach requires the solution of Hamilton-Jacoby-Bellman differential or different equation that is very difficult task because of well-known "dimension curse," if constraints are taken into account. Therefore, the most of the results presented in the literature are limited to the cases without constraints on the trading volume amounts. However, it's well-known that realistic investment models must include ones.

In this work, we consider the optimal portfolio selection problem under hard constraints on trading volume amounts when the dynamics of the risky asset returns are governed by a Markov-modulated Brownian motion. We assume that the mean return, variance and covariance of the risky assets switch over time according to a Markov chain, whose states are interpreted as the states of the economy. We propose to use the model predictive control (MPC) methodology in order to solve the problem. The major attraction of such technique lies in the fact that it can handle hard constraints on the inputs (manipulated variables) and states/outputs of a process and avoid the "curse of dimensionality" which, with the traditional dynamic programming-based approaches, hinders design of the feedback controls with constraints.

The main concept of MPC is to solve an open-loop constrained optimization problem at each time instant and implement only the initial optimizing control action of the solution. This procedure is repeated at the next time instant (see, for instance, Rawlings, 1999).

MPC have begun to be used with success in financial applications such as portfolio optimization and dynamic hedging. Some of the recent works on this subject can be found, for instance, in Dombrovskii et al (2005, 2006), Herzog et al. (2007), Primbs (2009), Primbs and Sung (2008). Dynamic hedging of basket options under soft probabilistic constraints using MPC is presented in Primbs (2009). Primbs and Sung (2008) develop stochastic receding horizon control for constrained index tracking problem in the presence of soft quadratic expectation constrains. Dombrovskii and others (2005, 2006) investigated control-theoretical approach to investment portfolio optimization with hard constraints using MPC. However, these papers do not account for jumps in stock returns. In Dombrovskii and Obyedko (2011) an approach to predictive control of systems with Markovian jumps and multiplicative noises was proposed. A simple example of its application to the investment portfolio optimization, based on simulation data, was presented.

In this paper, we consider the dynamic investment portfolio selection problem in the financial market with regime switching subject to hard constraints on trading volume amounts (a borrowing limit on the total wealth invested in the risky assets, and long- and shortsale restrictions on all risky assets). The problem is stated as a dynamic tracking problem of a reference portfolio with desired return under generalized performance criterion composed by a linear combination of quadratic and linear parts. The quadratic part represents the conditional mean-square error between the portfolio value and a reference portfolio and the liner part penalizes wealth values that less than the desired value. The open-loop feedback predictive trading strategy is derived based on MPC.

We pay a particular attention to testing of our approach on a set of a real data from radically different financial markets: the Russian Stock Exchange MICEX, the New York Stock Exchange and the Foreign Exchange Market (FOREX), taking into account the presence of transaction costs.

This work is organized as follows. Section 2 presents portfolio model and the optimization problem formulation. The main results of this article are presented in Section 3 where we design the optimal open-loop feedback control strategy for the problem under consideration. In Section 4 the numerical modeling results are presented. This paper is concluded in Section 5 with some final remarks.

## 2. Portfolio Model and Optimization Problem

Let us consider the investment portfolio of $n$ risky assets and one risk-free asset (e.g. a bank account or a government bond). Let $u_i(k)$, ($i=0,1,2,...,n$) denote the amount of money invested in the $i$th asset at time $k$; $u_0(k) \geq 0$ is the amount invested in a risk-free asset. Investor also can borrow the capital in case of need. The volume of the borrowing of the risk-free asset is equal to $u_{n+1}(k) \geq 0$. If $u_i(k)<0$, ($i=1,2,...,n$), then we use short position with the amount of shorting $|u_i(k)|$. The wealth process $V(k)$ satisfies

$$V(k) = \sum_{i=1}^{n} u_i(k) + u_0(k) - u_{n+1}(k). \qquad (1)$$

Let $P_i(k)$ denote the market value of the $i$th risky asset at time $k$, and $\eta_i(k+1)$ denote the corresponding return per period $[k,k+1]$, defined as

$$\eta_i(k+1) = \frac{P_i(k+1) - P_i(k)}{P_i(k)}.$$

It is a stochastic value unobserved at time $k$.

We consider self-financing portfolio. Self-financing means that the whole wealth obtained at the trading period $k$ will be exactly reinvested at the trading period $k+1$. By considering the self-finance strategies, the wealth dynamics are given by

$$V(k+1) = \sum_{i=1}^{n}[1+\eta_i(k+1)]u_i(k) + [1+r_1]u_0(k) - [1+r_2]u_{n+1}(k), \qquad (2)$$

with initial value $V(0)$, where $r_1$ is a risk-free interest rate, $r_2$ is a rate of borrowing of a risk-free asset ($r_1<r_2$).

Using (1), the dynamics in (2) can be rewritten as follows

$$V(k+1) = [1+r_1]V(k) + \sum_{i=1}^{n}[\eta_i(k+1)-r_1]u_i(k) - [r_2 - r_1]u_{n+1}(k), \qquad (3)$$

here $u_0(k) = V(k) - \sum_{i=1}^{n}u_i(k) + u_{n+1}(k)$ is the amount invested in a risk-free asset.

The returns of assets in which we are able to invest are described by a discrete-time approximation of the geometric Brownian motion with parameters depended on the state of the Markov chain $\alpha(k)$

$$\eta_i[\alpha(k),k] = \mu_i[\alpha(k),k] + \sum_{j=1}^{n}\sigma_{ij}[\alpha(k),k]w_j(k), \qquad (4)$$

where $\mu_i[\alpha(k),k]$ is the expected return of the $i$th risky asset; $\sigma[\alpha(k),k]=(\sigma_{ij}[\alpha(k),k])_{i,j=1,...,n}$ is the volatility matrix, $\sigma[\alpha(k),k]\sigma^T[\alpha(k),k]\geq 0$; the sequences $\{w_j(k); k=0,1,...; j=1,...,n\}$ are independent white noises with zero mean and unit variance; $\{\alpha(k); k=0,1,2,...\}$ is a finite-state discrete-time Markov chain taking values in $\{1,2,...,\nu\}$ with transition probability matrix

$$P = [P_{ij}], (i,j \in \{1,2,...,\nu\}),$$

$$P_{ij} = P\{\alpha(k+1) = \alpha_j | \alpha(k) = \alpha_i\}, \sum_{j=1}^{\nu}P_{ij} = 1,$$

and initial distribution

$$p_i = P\{\alpha(0) = i\}, (i=1,2,...,\nu); \sum_{i=1}^{\nu}p_i = 1.$$

We assume that $\alpha(k)$ and $w(k)$ are mutually independent. From (4), the expected return, variance and covariance of the assets at time $k$ are affected by local or global factors, which are represented by the market operation mode $\alpha(k)$. When the market operation mode is $\alpha(k)=\alpha_j$, then $\mu_i[\alpha(k),k] = \mu_i^{(j)}$, ($j=1,2...,\nu$; $i=1,2,...,n$) represents the expected return of the $i$th asset, while $\sigma[\alpha(k),k] = \sigma^{(j)}$ is the volatility matrix of the returns.

The Markovian chain defines the state of a market, e.g., a market in a state of high or low volatility and/or a market in a state of ascending or descending trend. We assume that at the instant of decision making, the current state of the market is known, i.e., the Markov state $\{\alpha(k)\}$ is observable. When practical problems are solved, indicators of a market state can be market indices.

Our objective is to control the investment portfolio, via dynamics asset allocation among the $n$ stocks and the risk-free asset, as closely as possible tracking the deterministic benchmark

$$V^0(k+1) = [1+\mu_0]V^0(k), \qquad (5)$$

where $\mu_0$ is a given parameter representing the growth factor, the initial state is $V^0(0)=V(0)$.

We impose the following constraints on the decision variables (a borrowing limit on the total wealth invested in the risky assets, and long- and short-sale restrictions on all risky assets)

$$u_i^{\min}(k) \leq u_i(k) \leq u_i^{\max}(k), (i=\overline{1,n}), \qquad (6)$$

$$0 \leq V(k) - \sum_{i=1}^{n}u_i(k) + u_{n+1}(k) \leq u_0^{\max}(k), \qquad (7)$$

$$0 \leq u_{n+1}(k) \leq u_{n+1}^{\max}(k). \qquad (8)$$

If $u_i^{\min}(k)<0$, ($i=1,2,...,n$), so we suppose that the amounts of the short-sale are restricted by $|u_i^{\min}(k)|$; if the short-selling is prohibited then $u_i^{\min}(k)\geq 0$, ($i=1,2,...,n$). The amounts of long-sale are restricted by $u_i^{\max}(k)$, ($i=1,2,...,n$); $u_0^{\max}(k)\geq 0$ defined the maximum amount of money we can invest in the risk-free asset; the borrowing amount is restricted by $u_{n+1}^{\max}(k) \geq 0$. Note, that values $u_i^{\min}(k)$, ($i=0,1,...,n$), $u_i^{\max}(k)$, ($i=0,1,...,n+1$) are often depend on common wealth of portfolio in practice. So that we can write $u_i^{\min}(k)=\beta_i V(k)$, $u_i^{\max}(k)=\gamma_i V(k)$, where $\beta_i$, $\gamma_i$ are constant parameters.

## 3. Model Predictive Control Strategy Design

We use the MPC methodology in order to define the optimal control portfolio strategy. The main concept of MPC is to solve an open-loop constrained optimization problem at each time instant and implement only the initial optimizing control action of the solution. Let evolution of a dynamic system be described by a discrete-time model and depend on the choice of controls $u(k)$. For the given prediction horizon $m$, a sequence of predictive controls $u(k/k), u(k+1/k),…,u(k+m−1/k)$ depending on the system state at the current time $k$ is calculated at each step $k$. This sequence optimizes the criterion chosen for the prediction horizon. At the time $k$, $u(k)=u(k/k)$ is assumed to be the control $u(k)$. Thereby, one gets control as a function of the current state, that is, feedback control. To obtain control at the next step $k+1$, the procedure is repeated, and the control horizon is one step shifted.

We consider the following quadratic objective with receding horizon (risk function)

$$J(k+m/k) = E\left\{\sum_{i=1}^{m}\left[V(k+i)-V^0(k+i)\right]^2 - \rho(k,i)\left[V(k+i)-V^0(k+i)\right]\right.$$
$$\left.+u^T(k+i-1/k)R(k,i-1)u(k+i-1/k) \Big/ V(k),\alpha(k)\right\}, \quad (9)$$

where $m$ is the prediction horizon, $u(k+i/k)=[u_1(k+i/k),…,u_{n+1}(k+i/k)]^T$ is the predictive control vector, $R(k)>0$ is a positive symmetric matrix of control cost coefficients, $\rho(k,i)>0$ is a positive weight coefficient; $E\{a/b\}$ is the conditional expectation of $a$ with respect to $b$. Notice that variable $V^0(k)$ is known for all time instant $k$ and may be considered as a pre-chosen parameter.

The performance criterion (9) is composed by a linear combination of a quadratic part, representing the conditional mean-square error between the investment portfolio value and a reference (benchmark) portfolio, and a linear part, penalizing wealth values that less than the desired value.

**Remark 1.** In the criterion (9), the terms involving quadratic forms on control vectors are added. In general, the presence of these terms, guarantees the existence of the control problem solution that will be shown below (see the remark 3 after Theorem 1).

**Remark 2.** Note that criterion (9) can be presented in the form

$$J(k+m/k) = \sum_{i=1}^{m} E\left\{\left[V(k+i)-E\{V(k+i)/V(k),\alpha(k)\}\right]^2 / V(k),\alpha(k)\right\}$$
$$+\left[E\{V(k+i)/V(k),\alpha(k)\}-V^0(k+i)\right]^2 \quad (10)$$
$$-\rho(k,i)\left[E\{V(k+i)/V(k),\alpha(k)\}-V^0(k+i)\right]$$
$$+E\{u^T(k+i-1/k)R(k,i-1)u(k+i-1/k)/V(k),\alpha(k)\}.$$

The performance criterion (10) is composed by a linear combination of the first term that is the conditional variance of the portfolio, the second term that is the quadratic error between the conditional mean of portfolio value and a reference portfolio, and a linear part, representing an error between the conditional mean portfolio value and a reference portfolio. If we eliminate the second term in (10), we obtain the multi-period conditional mean-variance criterion with receding horizon. In this case, the parameter $\rho(k,i)>0$ can be seen as a risk aversion coefficient, giving a trade-off between the expected portfolio value and the associated risk (variance) level.

An important advantage of tracking a reference portfolio approach under quadratic criterion (9) (or in its equivalent form (10)) is its capability to predict the trajectory of growth portfolio wealth, which would follow close to the deterministic (given by the investor) benchmark or beat it. It makes possible to obtain a smooth curve of the growth of the portfolio wealth on the entire investment horizon. It is one of the basic requirements for the trading strategies of investors in financial markets. The growth factor $\mu_0$ is selected by investor, based on the analysis of the financial market.

The discrete-time Markov chain, taking values in $\{1,2,…,v\}$, with transition probability matrix $P$ admits the following representation in the state space (Aggoun and Elliott, 2004)

$$\theta(k+1) = P\theta(k)+\upsilon(k+1), \quad (11)$$

where $\theta(k)=[\delta(\alpha(k),1), …,\delta(\alpha(k),v)]^T$, $\delta(\alpha(k),j)$ is a Kronecker function; $\{\upsilon(k)\}$ is a sequence of martingale increments with conditional moments

$$E\{\upsilon(k+1)/\theta(k)\}=0, \quad (12)$$

$$C(k+1) = E\{\upsilon(k+1)\upsilon^T(k+1)/\theta(k)\} = \text{diag}\{P\theta(k)\} - P\text{diag}\{\theta(k)\}P^T. \quad (13)$$

Taking (11) into consideration, equation (4) can be represented as follows

$$\eta_i[\theta(k),k] = \mu_i[\theta(k),k] + \sum_{j=1}^{n}\sigma_{ij}[\theta(k),k]w_j(k). \quad (14)$$

Criterion (9) can be transformed into

$$J(k+m/k) = E\left\{\sum_{i=1}^{m}\left[V(k+i)-V^0(k+i)\right]^2 - \rho(k,i)\left[V(k+i)-V^0(k+i)\right]\right. \tag{15}$$
$$\left. + u^T(k+i-1/k)R(k,i-1)u(k+i-1/k)\Big/V(k),\theta(k)\right\}.$$

The problem of minimizing the criterion (9) is equivalent to the quadratic control problem with criterion

$$J(k+m/k) = E\{\sum_{i=1}^{m} V^2(k+i) - R_1(k,i)V(k+i) + u^T(k+i-1/k)R(k,i-1)u(k+i-1/k)\Big/V(k),\theta(k)\}, \tag{16}$$

where we eliminated the term that is independent of control variables, $R_1(k+i) = 2V^0(k+i) + \rho(k,i)$.

We have the following theorem.

**Theorem 1.** Let the wealth dynamics is given by (3) with risky asset returns followed by dynamics of the form (14) under constraints (6)-(8). Then the MPC policy with receding horizon $m$, such that it minimizes the objective (15), for each instant $k$ is defined by the equation

$$u(k) = [I_{n+1} \quad 0_{n+1} \quad \ldots \quad 0_{n+1}]U(k), \tag{17}$$

where $I_{n+1}$ is $(n+1)$-dimensional identity matrix; $0_{n+1}$ is $(n+1)$-dimensional zero matrix; $U(k)=[u^T(k/k),\ldots, u^T(k+m-1/k)]^T$ is the set of predictive controls defined from the solving of quadratic programming problem with criterion

$$Y(k+m/k) = [2V(k)G(k) - F(k)]U(k) + U^T(k)H(k)U(k) \tag{18}$$

under constraints (element-wise inequality)

$$U_{\min}(k) \leq \overline{S}(k)U(k) \leq U_{\max}(k), \tag{19}$$

where

$$U_{\min}(k) = [u^T_{\min}(k), 0_{n+2\times 1}, \ldots, 0_{n+2\times 1}]^T, \quad U_{\max}(k) = [u^T_{\max}(k), 0_{n+2\times 1}, \ldots, 0_{n+2\times 1}]^T,$$

$$u_{\min}(k) = \begin{bmatrix} u_1^{\min}(k) \\ u_2^{\min}(k) \\ \ldots \\ u_n^{\min}(k) \\ -V(k) \\ 0 \end{bmatrix}, \quad u_{\max}(k) = \begin{bmatrix} u_1^{\max}(k) \\ u_2^{\max}(k) \\ \ldots \\ u_n^{\max}(k) \\ u_0^{\max}(k) - V(k) \\ u_{n+1}^{\max}(k) \end{bmatrix}.$$

$\overline{S}(k)$, $H(k)$, $G(k)$, $F(k)$ are the block matrices of the form

$$\overline{S}(k) = \text{diag}\{S(k), 0_{n+2\times n+1}, \ldots, 0_{n+2\times n+1}\},$$

$$H(k) = \begin{bmatrix} H_{11}(k) & H_{12}(k) & \ldots & H_{1m}(k) \\ H_{21}(k) & H_{22}(k) & \ldots & H_{2m}(k) \\ \ldots & \ldots & \ldots & \ldots \\ H_{m1}(k) & H_{m2}(k) & \ldots & H_{mm}(k) \end{bmatrix},$$

$$G(k) = [G_1(k) \quad G_2(k) \quad \ldots \quad G_m(k)],$$

$$F(k) = [F_1(k) \quad F_2(k) \quad \ldots \quad F_m(k)],$$

and the blocks satisfy the following recursive equations

$$S(k) = \begin{bmatrix} 1 & 0 & \ldots & 0 & 0 \\ 0 & 1 & \ldots & 0 & 0 \\ \ldots & \ldots & \ldots & \ldots & \ldots \\ 0 & 0 & \ldots & 1 & 0 \\ -1 & -1 & \ldots & -1 & 1 \\ 0 & 0 & \ldots & 0 & 1 \end{bmatrix},$$

$$H_{tt}(k) = R(k,t-1) + Q_1(m-t)\sum_{q=1}^{\nu}\left[e_q \text{diag}\{P^t\theta(k)\}e_q^T\right]\sum_{j=0}^{n}\left(B_j^{(q)}(k+t)\right)^T B_j^{(q)}(k+t), \tag{20}$$

$$H_{tf}(k) = A^{f-t}Q_1(m-f)\sum_{q=1}^{\nu}\sum_{r=1}^{\nu}\left[e_r \text{diag}\{P^f\theta(k)\}\left(P^{f-t}\right)^T e_q^T\right]\left(B_0^{(q)}(k+t)\right)^T B_0^{(r)}(k+f), t < f, \tag{21}$$

$$H_{tf}(k) = H_{ft}^T(k), t > f, \quad (t,f = \overline{1,m}), \tag{22}$$

$$G_t(k) = A^t Q_1(m-t)\sum_{q=1}^{\nu}\left[e_q P^t\theta(k)\right]B_0^{(q)}(k+t), \tag{23}$$

$$F_t(k) = Q_2(m-t)\sum_{q=1}^{\nu}\left[e_q P^t \theta(k)\right] B_0^{(q)}(k+t), \quad (24)$$

$Q_1(t) = A^2 Q_1(t-1) + 1, \ Q_1(0) = 1,$

$Q_2(t) = A Q_2(t-1) + R_1(k, m-t), \ Q_2(0) = R_1(k,m),$

$R_1(k,t) = 2V^0(k+t) + \rho(k,t), \ (t = \overline{1,m}),$

$A = 1 + r_1,$

$e_q = [0,...,0,1,0,...,0]_{1\times\nu}, (q = \overline{1,\nu}),$

$B_0^{(q)}(k) = \left[\mu_1^{(q)} - r_1 \quad \mu_2^{(q)} - r_1 \quad ... \quad \mu_n^{(q)} - r_1 \quad r_1 - r_2\right],$

$B_j^{(q)}(k) = \left[\sigma_{1j}^{(q)} \quad ... \quad \sigma_{nj}^{(q)} \quad 0\right], (j = \overline{1,n}), (q = \overline{1,\nu}).$

A proof of this theorem is reported in the Appendix A.

**Remark 3.** The condition $R(k,i) > 0$ guarantees that the matrix $H(k)$ is positive definite, and thus the solution of a quadratic programming problem with a criterion (18) exists and is unique, if the constraints (19) are admissible.

## 4. Numerical Experiments and Discussion

In this section, we present several numerical examples demonstrating the application of our approach to portfolios of a real stocks and currency pairs. We want to assess the performance of our model under real market conditions by computing the portfolio wealth over a long period of time. The data used for these examples are taken from the Russian Stock Exchange MICEX, the New York Stock Exchange and the Foreign Exchange Market (FOREX) (www.finam.ru). They include the daily stock prices of the largest companies, currency pairs as well as the values of the MICEX index and the Dow Jones Index.

We consider the situation of an investor who has to allocate one unit of wealth over the investment horizon of about 600 trading days among risky assets and one risk-free asset. The updating of the portfolio is executed once every trading day.

The wealth of the real portfolio was evaluated for the case when each portfolio rebalancing brings the transaction costs. Let the investor pays fraction $c_i > 0, (i=1,...,n)$ of the amount transacted on purchase and sale of the $i$th stock. The total cost due to all transactions at time $k$ is then $\sum_{i=1}^{n} c_i |u_i(k) - u_i(k-1)|$, where $|u_i(k) - u_i(k-1)|$ is the amount of money transacted on purchase and sale of the $i$th asset per period $[k-1,k]$. In presence of transaction costs, a self-financing portfolio strategy does not allow additional money to be taken into the portfolio to cover the trading costs. These costs should thus be paid from the invested wealth.

Taking transaction costs into consideration, the wealth of the real portfolio at time $k+1$ was defined as

$$V(k+1) = [1+r_1]V(k) + \sum_{i=1}^{n}[\eta_i(k+1) - r_1]u_i(k) - \sum_{i=1}^{n} c_i |u_i(k) - u_i(k-1)| - [r_2 - r_1]u_{n+1}(k). \quad (25)$$

So, we evaluate the wealth of the real portfolio taking into account the transaction costs. We use portfolio value computed by (25) in equation (18). This approach allows us to effectively avoid non-convex optimization problem that usually arises while dealing with portfolio optimization subject to transaction costs.

The risk-free asset is considered here as a bank account. For the sake of simplicity, we assume that $r_2 = r_1 = 0$. We also don't take into account cross-sectional correlation between different assets, i.e., $\sigma_{ij} = 0$, $i \neq j$. We have experimented with more sophisticated scheme, under assumption that cross-correlation between assets is presented. However, we found its impact on the tracking performance quite negative. This is expected, since we need to estimate a large number of parameters that introduces "estimation uncertainty" into the portfolio optimization strategy.

In the first set of experiments, we used risky assets traded on the Russian Stock Exchange MICEX: Sberbank, Gazprom, VTB, LUKOIL, NorNickel, Rosneft, Gazpromneft. All investment portfolios were composed of 5 risky assets. The relatively small size of the Russian Stock Exchange precluded experiments with a large number of assets. On the over hand, it presented additional challenges due to the high volatility of the index.

We assume that the market parameters depend on the market mode that switches according to a Markov chain among two states. We use MICEX Index to observe the current Markov state and to estimate the transition probability matrix. We assume that only volatilities of the returns are effected by local or global factors, which are represented by the market regime. Thus, State 1 represents low market volatility and State 2 represents high market volatility. Whenever the daily volatility of the index was below 0.015, we defined that day as low-volatility and set $\sigma_{ii}^{(1)} = 0.01$, $(i=1,2,...,n)$. Whereas the daily volatility of the index was above 0.015, we defined the day as a high-volatility and set $\sigma_{ii}^{(2)} = 0.02$, $(i=1,2,...,n)$. These values of volatilities were obtained by analyzing a real market behavior.

The transition probability matrix was estimated by the maximum likelihood method using the past 200 daily closing values of the MICEX index prior to the tracking period. The estimation of transition probability matrix was

$$P = \begin{bmatrix} 0.96 & 0.24 \\ 0.04 & 0.76 \end{bmatrix}.$$

The Markov process is assumed to be a stationary multi-period process over the investment horizon.

We computed the expected returns using 13-day simple averaging of past historical return data and assume that the expected returns remain constant over the predictive horizon $m$. We use the adjusted procedure, updating the estimates at each decision time $k$, to adapt the portfolio to price changes on the market incorporating of newly arrived information.

We set the tracking target to return 0.15% per day ($\mu_0=0.0015$). We assumed an initial portfolio wealth of $V(0)=V^0(0)=1$. The weight coefficients are set as $R(k,i)=\text{diag}(10^{-4},\dots,10^{-4})$, $\rho(k,i)=0.1$ for all $k,i$. We suppose that the fractions $c_i$, ($i=1,\dots,n$) are equal to 0.0006 (0.06 %). We impose constraints on the tracking portfolio problem with parameters $\beta_i=-0.6$, ($i=1,\dots,n$), $\gamma_i=3$, ($i=1,\dots,n+1$). Therefore, in all examples, we allow borrowing and short selling. A prediction horizon was $m=10$. The optimization problem (18), (19) is solved by the quadprog.m function in MATLAB Optimization Toolbox.

We present the typical results of the experiments on Fig. 1-3. In the pictures below the portfolio was composed of five risky assets: LUKOIL, Gazprom, Sberbank, Sibneft, NorNickel. The investment period was from August 10, 2011 to October 20, 2013. Fig. 1 plots real portfolio and benchmark values. In Fig. 2 we have investments in the risky asset Gazprom. Fig. 3 illustrates the MICEX Index daily returns and the estimated states of the Markov chain.

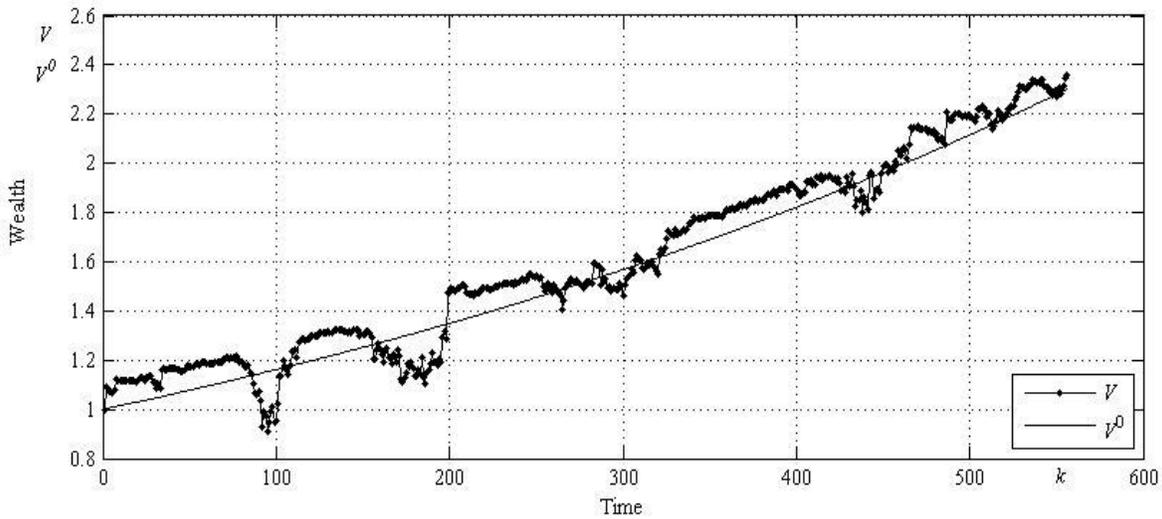

**Fig 1.** Tracking performance ($V$ – real portfolio, $V^0$ – reference portfolio).

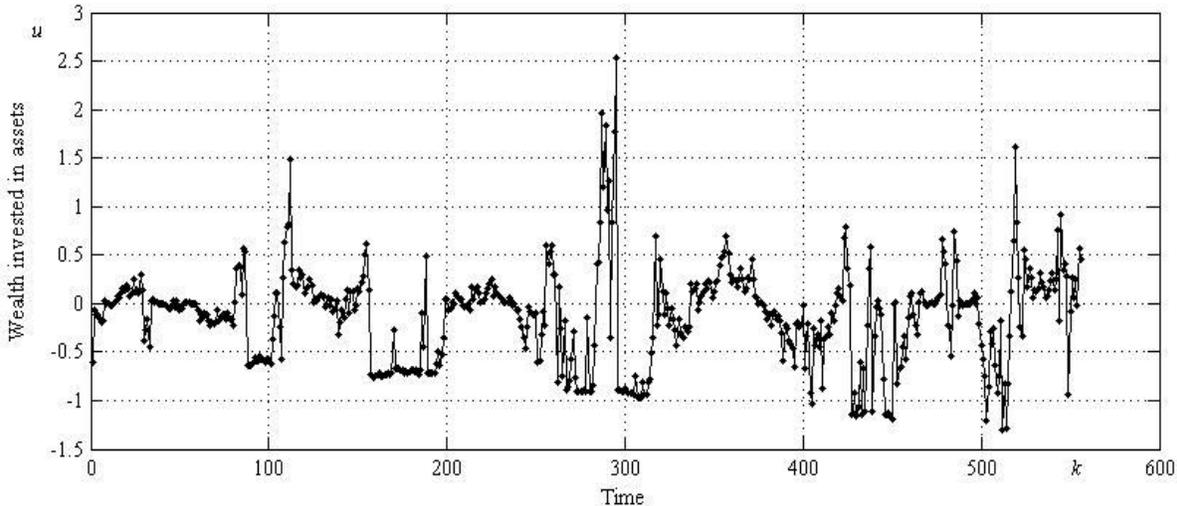

**Fig.2.** Asset allocation decision ($u$ is the amount invested in Gazprom).

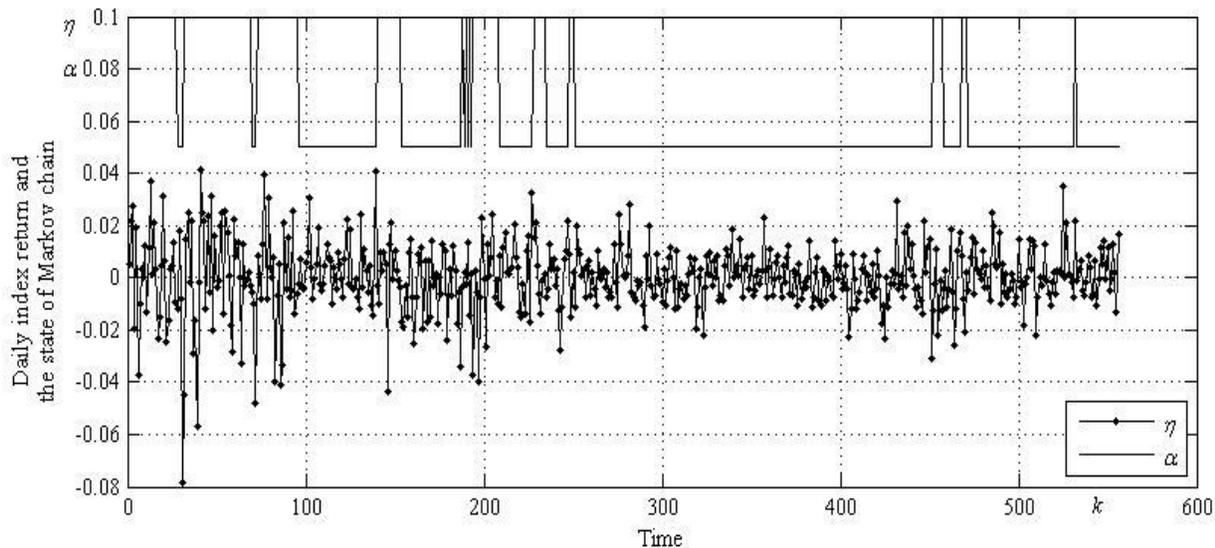

**Fig. 3.** Daily MICEX return and the state of Markov chain.

In the second set of experiments, we used risky assets traded on the New York Stock Exchange: AT&T Inc, Apple Inc, Bank of America, Caterpillar Inc, Cisco Systems Inc, Citigroup Inc, Google Inc, IBM, Intel Corp, JPMorgan Chase&Co, Microsoft Corp, United Technologies corp. All investment portfolios were composed of 6 risky assets.

We use Dow Jones Index to observe the current Markov state and to estimate the transition probability matrix. Whenever the daily volatility of the index was below 0.01, we defined that day as low-volatility and set $\sigma_{ii}^{(1)}=0.005$ ($i=1,2,…,n$). Whereas the daily volatility of the index was above 0.01, we defined the day as a high-volatility and set $\sigma_{ii}^{(2)}=0.02$ ($i=1,2,…,n$). These values of volatility were obtained by analyzing a real American market behavior. It should be noted that the volatility of the American market is lower than one on the Russian market. The estimation of transition probability matrix was

$$P = \begin{bmatrix} 0.91 & 0.15 \\ 0.09 & 0.85 \end{bmatrix}.$$

The expected returns were estimated using 21-day simple averaging. We set the tracking target to return 0.1% per day ($\mu_0=0.001$) and $\rho=0.1$. The other parameters were the same as in the first example.

We present the typical results of the experiments on Fig. 4-6. In the pictures below the portfolio was composed of risky assets: Apple Inc, Cisco Systems Inc, Caterpillar Inc, Google Inc, Intel Corp, United Technologies corp. The investment period was from June 07, 2011 to October 20, 2013.

Fig. 4 plots the real portfolio values and the reference portfolio values. In Fig. 5 we have investments in the risky assets Apple Inc. Fig. 6 illustrates the Dow Jones Index daily returns and the estimated states of the Markov chain.

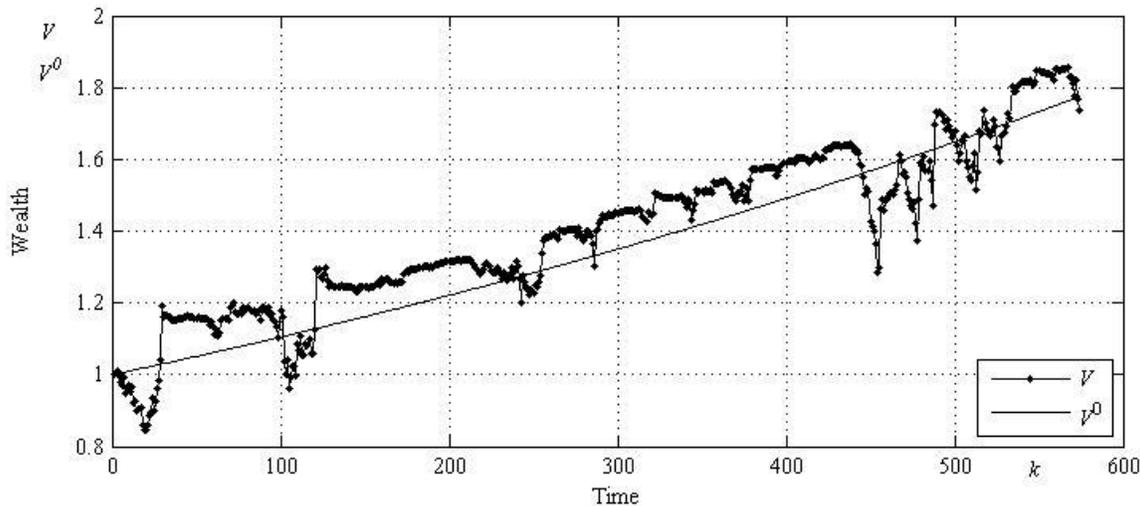

**Fig 4.** Tracking performance ($V$ – real portfolio, $V^0$ – reference portfolio).

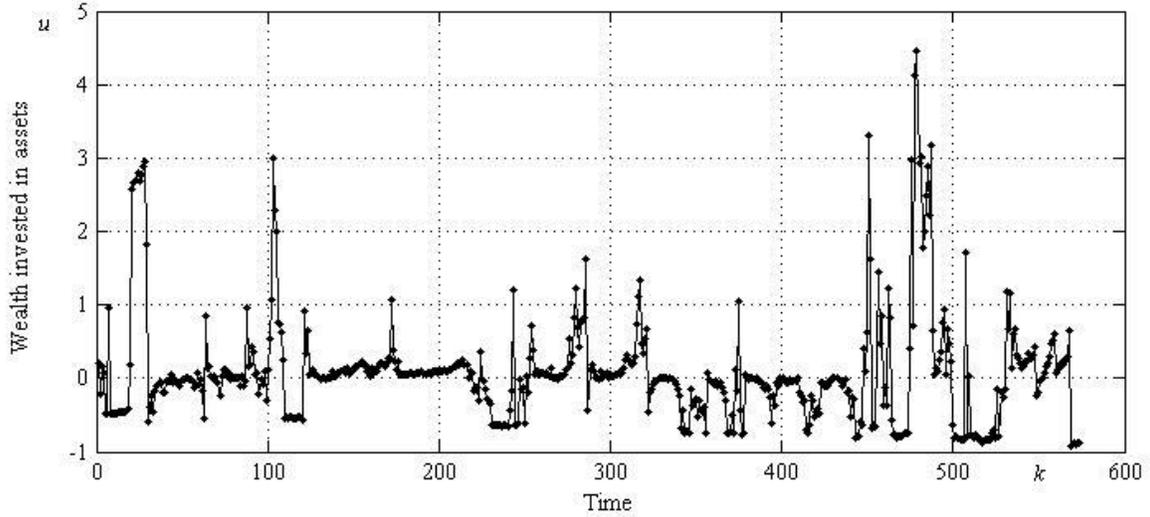

**Fig.5.** Asset allocation decision ($u$ – is the amount invested in Apple Inc).

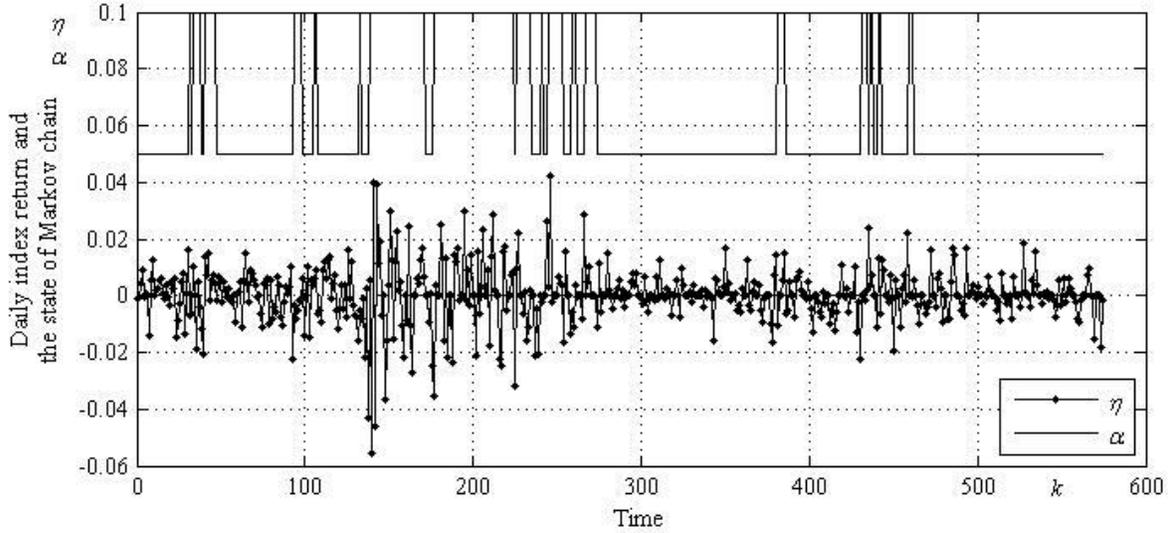

**Fig. 6.** Daily D&J return and the state of Markov chain.

In the third set of experiments we used currency pairs trading on the Foreign Exchange Market (FOREX): USD/JPY, EUR/JPY, EUR/USD, USD/CHF, USD/DEM, EUR/GBP, GBP/USD. Currencies are traded against one another. Each currency pair thus constitutes an individual trading product. All investment portfolios were composed of 5 currency pairs.

We use pair EUR/USD to observe the current Markov state and to estimate the transition probability matrix. Whenever the daily volatility of the index was below 0.006, we defined that day as low-volatility and set $\sigma_{ii}^{(1)}$=0.004 ($i$=1,2,…,$n$). Whereas the daily volatility of the index was above 0.006, we defined the day as a high-volatility and set $\sigma_{ii}^{(2)}$=0.008 ($i$=1,2,…,$n$). These values of volatility were obtained by analyzing a real currency market behavior. The estimation of transition probability matrix was

$$P = \begin{bmatrix} 0.91 & 0.16 \\ 0.09 & 0.84 \end{bmatrix}.$$

The expected returns were estimated using 21-day simple averaging. We impose constraints on the tracking portfolio problem with parameters $\beta_i$=-1, ($i$=1,…,$n$), $\gamma_i$=5, ($i$=1,…,$n$+1). Note that the foreign exchange market is unique because of the low margins of relative profit and the use of high leverage to enhance profit and loss margins and with respect to account size. We suppose that the fractions $c_i$, ($i$=1,…,$n$) are equal to 0.0001 (0.01 %) and $\rho$=0.2. We set the tracking target to return 0.15% per day ($\mu_0$=0.0015) and $\rho$=0.1. The other parameters were the same as in the first example.

We present the typical results of the experiments on Fig. 7-9. In the pictures below the portfolio was composed of currency pairs: EUR/GBP, EUR/USD, USD/CHF, USD/JPY, USD/DEM. The investment period was from February 22, 2012 to October 20, 2013.

Fig. 7 plots the real portfolio values and the reference portfolio values. In Fig. 8 we have investments in the currency EUR/USD. Fig. 9 illustrates the EUR/USD daily returns and the estimated states of the Markov chain.

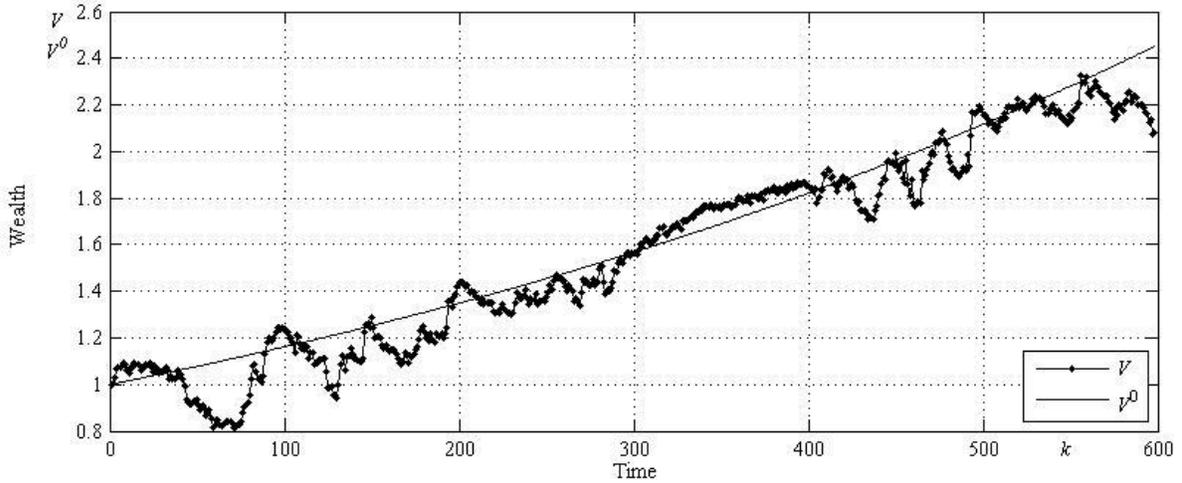

**Fig.7.** Tracking performance ($V$ – real portfolio, $V^0$ – reference portfolio).

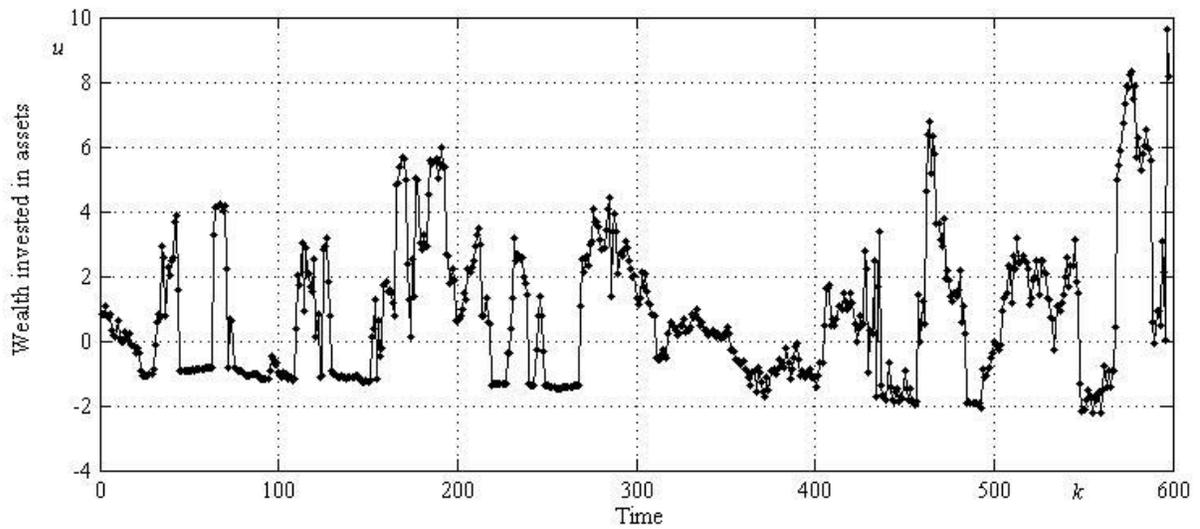

**Fig.8.** Asset allocation decision ($u$ is the amount invested in EUR/USD).

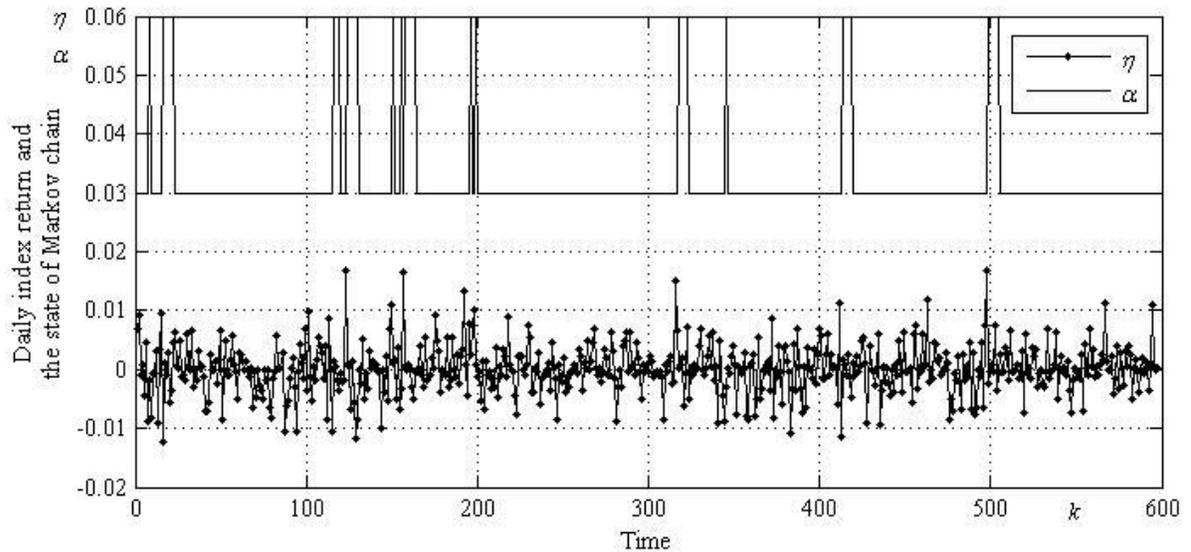

**Fig. 9.** Daily EUR/USD return and the state of Markov chain.

Several insights can be garnered from the examples illustrated above. It is important to acknowledge that in our experiments where we use rather simple methods for parameters estimation, the tracking performance appears to be rather efficient. One of the major attractions of proposed MPC algorithm lies in the fact that it appears to be rather insensitive to the estimated parameters that are fed into the model. Our approach does not require a heavy reliance on parametric estimation based on past data; instead it focuses on trying to capture the dynamic changes of the market over the tracking period and react accordingly. So our approach allows us to design strategies which are desensitized, i.e., robustified, to parameters' estimation. It should, however, be realized that the results of an ensuing optimization model will always be affected by a level of "estimation uncertainty." It is clear that one can use more sophisticated estimation schemes to improve the parameters estimation precision (Elliott, Malcolm and Tsoi, 2003; Sims, Waggoner and Zha, 2008).

## 5. Conclusion

In this paper, we studied a discrete-time portfolio selection problem subject to Markovian jumps in the parameters. We proposed to use the MPC methodology in order to solve the problem. The optimal open-loop feedback portfolio control strategy was derived under hard constraints on trading volume amounts. The advantage of using a receding horizon implementation is that at each decision stage we can profit from observations of actual market behavior during the preceding period and use information to feed fresh estimates to the model.

We presented the numerical modeling results, based on a set of real data from the Russian Stock Exchange MICEX, the New York Stock Exchange and the Foreign Exchange Market (FOREX). We find that on actual data the proposed approach is reasonable. The value of the portfolio follows the value of the reverence portfolio, beating it most of the time and the constraints are satisfied.

The main features of our approach are (a) the ability to adapt to switching market environments (regimes) by dynamically incorporating new information into the decision process; (b) the flexibility of dealing with portfolio constraints and transaction costs; (c) the results of optimization are rather insensitive to errors in the estimation.

Possible future work will focus on generalizing the results to more sophisticated asset returns models, for example, autoregressive conditional heteroskedasticity models with regime switching (Hamilton and Susmel, 1994) or structural vector autoregressions with Markov switching (Lanne, Lütkepohl and Maciejowska, 2010).

## Appendix A. Proof of the Theorem 1.

The portfolio dynamics (3) can be rewritten in the form

$$V(k+1) = AV(k) + B_0[\theta(k+1), k+1]u(k) + \sum_{j=1}^{n} B_j[\theta(k+1), k+1]w_j(k+1)u(k), \quad \text{(A.1)}$$

where

$$u(k) = [u_1(k) \quad \ldots \quad u_{n+1}(k)]^T,$$
$$A = 1 + r_1,$$
$$B_0[\theta(k), k] = [\mu_1[\theta(k), k] - r_1 \quad \ldots \quad \mu_n[\theta(k), k] - r_1 \quad r_1 - r_2],$$
$$B_j[\theta(k), k] = [\sigma_{1j}[\theta(k), k] \quad \ldots \quad \sigma_{nj}[\theta(k), k] \quad 0], (j = \overline{1, n}),$$
$$B_j[\theta(k), k] = \sum_{q=1}^{v} \theta_q(k) B_j^{(q)}(k), (j = \overline{0, n}). \quad \text{(A.2)}$$

Here $\theta_i(k)$, ($i=1,2,\ldots,v$) are elements of the vector $\theta(k)$; $\{B_j^{(q)}(k)\}$, ($j=0,\ldots,n$), ($q=1,\ldots,v$) is a set of values of the matrix $B_j[\theta(k),k)]$.

The constraints (6)-(8) can be rewritten in matrix form (element-wise inequality)

$$u_{\min}(k) \leq S(k)u(k) \leq u_{\max}(k), \quad \text{(A.3)}$$

where

$$S(k) = \begin{bmatrix} 1 & 0 & \ldots & 0 & 0 \\ 0 & 1 & \ldots & 0 & 0 \\ 0 & 0 & \ldots & 0 & 0 \\ 0 & 0 & \ldots & 1 & 0 \\ -1 & -1 & \ldots & -1 & 1 \\ 0 & 0 & \ldots & 0 & 1 \end{bmatrix}.$$

Thus, we have a control problem to minimize a functional (15), with system dynamics like (A.1) under constraints (A.3).

The objective (15) can be written in the form

$$J(k+m/k) = E\{X^T(k+1)X(k+1) - \Delta_1(k+1)X(k+1) + U^T(k)\Delta(k)U(k)/V(k), \theta(k)\}, \quad \text{(A.4)}$$

subject to
$$X(k+1) = \Psi V(k) + \Phi_0[\Xi(k+1),k+1]U(k) + \sum_{j=1}^{n}\Phi_j[\Xi(k+1),k+1]\text{diag}\{W_j(k+1)\}U(k), \quad (A.5)$$
where
$$X(k+1) = \begin{bmatrix} V(k+1) \\ V(k+2) \\ \dots \\ V(k+m) \end{bmatrix}, \Psi = \begin{bmatrix} A \\ A^2 \\ \dots \\ A^m \end{bmatrix}, U(k) = \begin{bmatrix} u(k/k) \\ u(k+1/k) \\ \dots \\ u(k+m-1/k) \end{bmatrix}, \Xi(k+1) = \begin{bmatrix} \theta(k+1) \\ \theta(k+2) \\ \dots \\ \theta(k+m) \end{bmatrix}, W_j(k+1) = \begin{bmatrix} w_j(k+1) \\ w_j(k+2) \\ \dots \\ w_j(k+m) \end{bmatrix},$$

$$\Phi_j[\Xi(k+1),k+1] = \begin{bmatrix} B_j[\theta(k+1),k+1] & 0_{1\times n+1} & \dots & 0_{1\times n+1} \\ AB_j[\theta(k+1),k+1] & B_j[\theta(k+2),k+2] & \dots & 0_{1\times n+1} \\ \dots & \dots & \dots & \dots \\ A^{m-1}B_j[\theta(k+1),k+1] & A^{m-2}B_j[\theta(k+2),k+2] & \dots & B_j[\theta(k+m),k+m] \end{bmatrix}, (j=0,1,\dots,n),$$

$$\Delta(k) = \text{diag}\{R(k,0),R(k,1),\dots,R(k,m-1)\}, \quad \Delta_1(k+1) = [R_1(k,1),R_1(k,2),\dots,R_1(k,m)].$$

Using (A.5), we can rewrite (A.4) as follows
$$\begin{aligned}J(k+m/k) &= V^2(k)\Psi^T\Psi - \Delta_1\Psi V(k) \\
&+ \left[2V(k)\Psi^T - \Delta_1(k+1)\right]E\{\Phi_0[\Xi(k+1),k+1]/\theta(k)\}U(k) \\
&+ U^T(k)\Big[E\{\Phi_0^T[\Xi(k+1),k+1]\Phi_0[\Xi(k+1),k+1] \\
&+ \sum_{j=1}^{n}\text{diag}\{W_j(k+1)\}\Phi_j^T[\Xi(k+1),k+1]\Phi_j[\Xi(k+1),k+1]\text{diag}\{W_j(k+1)\}/\theta(k)\} + \Delta(k)\Big]U(k).\end{aligned} \quad (A.6)$$

Denote the matrices
$$H(k) = E\{\Phi_0^T[\Xi(k+1),k+1]\Phi_0[\Xi(k+1),k+1] + \sum_{j=1}^{n}\text{diag}\{W_j(k+1)\}\Phi_j^T[\Xi(k+1),k+1]$$
$$\cdot \Phi_j[\Xi(k+1),k+1]\text{diag}\{W_j(k+1)\}/\theta(k)\} + \Delta(k),$$
$$G(k) = \Psi^T E\{\Phi_0[\Xi(k+1),k+1]/\theta(k)\},$$
$$F(k) = \Delta_1(k+1)E\{\Phi_0[\Xi(k+1),k+1]/\theta(k)\}.$$

Notice that the matrices $H(k)$, $G(k)$, $F(k)$ don't depend on $V(k)$.

We have that the minimization of the criterion (A.4) under constraints (A.3) is equivalent to the quadratic programming problem with criterion (18)
$$Y(k+m/k) = [2V(k)G(k) - F(k)]U(k) + U^T(k)H(k)U(k)$$

under constraints (19). Straightforward calculations lead to the following expression for the blocks $\{H_{tf}(k)\}$, $(t,f=\overline{1,m})$, of the matrix $H(k)$

$$H_{tt}(k) = E\{\sum_{i=t}^{m}A^{2(i-t)}B_0^T[\theta(k+t),k+t]B_0[\theta(k+t),k+t]$$
$$+ \sum_{i=t}^{m}A^{2(i-t)}\sum_{j=1}^{n}B_j^T[\theta(k+t),k+t]B_j[\theta(k+t),k+t]w_j^2(k+t)/\theta(k)\} + R(k,t-1),$$

$$H_{tf}(k) = E\{\sum_{i=f}^{m}A^{2(i-f)}A^{f-t}B_0^T[\theta(k+t),k+t]B_0[\theta(k+f),k+f]$$
$$+ \sum_{i=f}^{m}A^{2(i-f)}A^{f-t}\sum_{j=1}^{n}\sum_{l=1}^{n}B_j^T[\theta(k+t),k+t]B_l[\theta(k+f),k+f]w_j(k+t)w_l(k+f)/\theta(k)\}, t<f,$$

$$H_{ft}(k) = H_{tf}, t>f.$$

Let introduce the following recursive equation
$$Q_1(t) = A^2 Q_1(t-1) + 1, (t=\overline{1,m}),$$
starting with $Q_1(0)=1$. Then using the fact that the sequences $\{w_j(k); k=0,1,\dots; j=1,\dots,n\}$ are independent white noises with zero mean and unity variance, the blocks $\{H_{tf}(k)\}$ $(t,f=\overline{1,m})$, can be written as

$$H_{tt}(k) = Q_1(m-t)\sum_{j=0}^{n} E\left\{B_j^T[\theta(k+t),k+t]B_j[\theta(k+t),k+t]/\theta(k)\right\} + R(k,t-1),$$

$$H_{tf}(k) = A^{f-t}Q_1(m-f)E\left\{B_0^T[\theta(k+t),k+t]B_0[\theta(k+f),k+f]/\theta(k)\right\}, t < f.$$

From (11) and (A.2), the following expressions hold

$$\theta(k+t) = P^t\theta(k) + \sum_{l=1}^{t} P^{t-l}\upsilon(k+l),$$

$$B_j[\theta(k+t),k+t] = \sum_{q=1}^{v}\theta_q(k+t)B_j^{(q)}(k+t) = \left[\sum_{q=1}^{v}e_q P^t\theta(k) + \sum_{q=1}^{v}e_q\sum_{l=1}^{t}P^{t-l}\upsilon(k+l)\right]B_j^{(q)}(k+t), j = \overline{0,n}, \quad (A.7)$$

where $e_q = [0,...,0,1,0,...,0]_{1\times v}, (q = \overline{1,v})$.

Based on the above, we have

$$H_{tt}(k) = Q_1(m-t)\sum_{j=0}^{n}\sum_{q=1}^{v}\sum_{r=1}^{v}e_r\left\{P^t\theta(k)\theta^T(k)(P^t)^T + \sum_{l=1}^{t}P^{t-l}C(k+l)(P^{t-l})^T\right\}e_q^T(B_j^{(q)}(k+t))^T B_j^{(r)}(k+t) + R(k,t-1),$$

$$H_{tf}(k) = A^{f-t}Q_1(m-f)\sum_{q=1}^{v}\sum_{r=1}^{v}e_r\left\{P^f\theta(k)\theta^T(k)(P^t)^T + \sum_{l=1}^{t}P^{f-l}C(k+l)(P^{t-l})^T\right\}e_q^T(B_0^{(q)}(k+t))^T B_0^{(r)}(k+f),$$

where $C(k+l) = E\{\upsilon(k+l)\upsilon^T(k+l)/\theta(k)\}$.

From the definition of $\theta(k)$, we have

$$\theta(k+t)\theta^T(k+t) = \text{diag}\{\theta(k+t)\}. \quad (A.8)$$

It is easy to show that

$$C(k+l) = E\{\upsilon(k+l)\upsilon^T(k+l)/\theta(k)\} = \text{diag}\{P^l\theta(k)\} - P\text{diag}\{P^{l-1}\theta(k)\}P^T. \quad (A.9)$$

Using expressions (A.8) and (A.9), we get after some algebraic manipulations that

$$H_{tt}(k) = Q_1(m-t)\sum_{j=0}^{n}\sum_{q=1}^{v}\sum_{r=1}^{v}e_r\left\{P^t\text{diag}\{\theta(k)\}(P^t)^T + \sum_{l=1}^{t}P^{t-l}\text{diag}\{P^l\theta(k)\}(P^{t-l})^T\right.$$

$$\left.-\sum_{l=1}^{t}P^{t-l}P\text{diag}\{P^{l-1}\theta(k)\}P^T(P^{t-l})^T\right\}e_q^T(B_j^{(q)}(k+t))^T B_j^{(r)}(k+t) + R(k,t-1)$$

$$= Q_1(m-t)\sum_{j=0}^{n}\sum_{q=1}^{v}\left[e_q\text{diag}\{P^t\theta(k)\}e_q^T\right](B_j^{(q)}(k+t))^T B_j^{(q)}(k+t) + R(k,t-1),$$

$$H_{tf}(k) = A^{f-t}Q_1(m-f)\sum_{q=1}^{v}\sum_{r=1}^{v}\left[e_r\text{diag}\{P^f\theta(k)\}\left(P^{f-t}\right)^T e_q^T\right](B_0^{(q)}(k+t))^T B_0^{(r)}(k+f).$$

It is easy to show that the blocks of the matrices $G(k), F(k)$ are of the form

$$G_t(k) = E\left\{\sum_{i=t}^{m}A^i A^{i-t}B_0[\theta(k+t),k+t]/\theta(k)\right\} = A^t Q_1(m-t)E\left\{B_0[\theta(k+t),k+t]/\theta(k)\right\},$$

$$F_t(k) = E\left\{\sum_{i=t}^{m}R_1(k,i)B_0[\theta(k+t),k+t]/\theta(k)\right\} = Q_2(m-t)E\left\{B_0[\theta(k+t),k+t]/\theta(k)\right\},$$

where

$$Q_2(t) = AQ_2(t-1) + R_1(k,m-t), Q_2(0) = R_1(k,m), (t = \overline{1,m}).$$

Using (A.7), we have

$$G_t(k) = A^t Q_1(m-t)\sum_{q=1}^{v}\left[e_q P^t\theta(k)\right]B_0^{(q)}(k+t),$$

$$F_t(k) = Q_2(m-t)\sum_{q=1}^{v}\left[e_q P^t\theta(k)\right]B_0^{(q)}(k+t).$$

We have that the blocks of the matrices $H(k), G(k), F(k)$ take the forms (20)-(24). This completes the proof.

**References**

Aggoun, L., Elliott, R.J., 2004. Measure theory and filtering. Cambridge University Press, New York.

Bäuerle, N., Rieder, U., 2004. Portfolio optimization with Markov-modulated stock prices and interest rates. IEEE Transactions on Automatic Control 49 (3), 442-447.

Billio, M., Pelizzon, L., 2000. Value-at-Risk: a multivariate switching regime approach. J. of Empirical Finance 7, 531-554.

Cogley, T., Sargent, T.J., 2009. Diverse beliefs survival and the market price of risk. Economic Journal 119 (536), 354-376.

Costa, O.L.V., Araujo, M.V., 2008. A generalized multi-period portfolio optimization with Markov switching parameters. Automatica 44 (10), 2487-2497.

Dombrovskii, V.V., Dombrovskii D.V., Lyashenko, E.A., 2005. Predictive control of random-parameter systems with multiplicative noise. Application to investment portfolio optimization. Automation and remote control 66 (4), 583-595.

Dombrovskii, V.V., Dombrovskii D.V., Lyashenko, E.A., 2006. Model predictive control of systems with random dependent parameter under constraints and It's application to the investment portfolio optimization. Automation and remote control 67 (12), 1927-1939.

Dombrovskii, V.V., Ob"edko, T.Yu., 2011. Predictive control of systems with Markovian jumps under constraints and it's application to the investment portfolio optimization. Automation and remote control 72 (5), 989-1003.

Elliott, R.J., Van der Hoek, J., 1997. An application of hidden Markov models to asset allocation problems. Finance and Stochastics 1, 229-238.

Elliott, R.J., Malcolm, W.P., Tsoi, A.H., 2003. Robust parameter estimation for asset price models with Markov modulated volatilities. Journal of Economic Dynamics and Control 27, 1391-1409.

Hamilton, J.D., Susmel, R., 1994. Autoregressive conditional heteroskedasticity and changes in regime. Journal of Econometrics 64, 307-333.

Hansen, L.P., Mayer, R., Sargent, T., 2010. Robust hidden Markov LQG problems. Journal of Economic Dynamics and Control 34. 1951-1966.

Herzog, F., Dondi, G., Geering, H.P., 2007. Stochastic model predictive control and portfolio optimization. International Journal of Theoretical and Applied Finance 10 (2), 203-233.

Honda, T., 2003. Optimal portfolio choice for unobservable and regime-switching mean returns. Journal of Economic Dynamics and Control 28, 45-78.

Lanne, M., Lütkepohl, H., Maciejowska, K., 2010. Structural vector autoregressions with Markov switching. Journal of Economic Dynamics and Control 34, 121-131.

Liu, H., 2011. Dynamic portfolio choice under ambiguity and regime switching mean returns. Journal of Economic Dynamics and Control 35, 623-640.

Marcowitz, H.M., 1952.Portfolio selection. J. Finance 7 (1), 77-91.

Merton, R.C., 1990. Continuous-time finance. Cambridge: Balckwell.

Galperin, V.A., Dombrovskii, V.V., Fedosov, E.N., 2005. Dynamic control of the investment portfolio in the jump-diffusion financial market with regime-switching. Automation and Remote Control 66 (5), 837-850.

Gerasimov, E.S., Dombrovskii, V.V., 2003. Dynamic network model of managing investment portfolio under random stepwise changes in volatilities of financial assets. Automation and Remote Control 64 (7), 1086-1099.

Primbs, J.A., 2009. Dynamic hedging of basket options under proportional transaction costs using receding horizon control. Int. J. of Control 82 (10), 1841-1855.

Primbs, J.A., Sung, C.H., 2008. A stochastic receding horizon control approach to constrained index tracking. Asia-Pacific Finan Markets 15, 3-24.

Rawlings, J., 1999. Tutorial: Model Predictive Control Technology. In Proc. Amer. Control Conf., 662-676.

Sims, A.C., Waggoner, D.F., Zha, T., 2008. Methods for inference in large multiple-equation Markov-switching models. Journal of Econometrics 146. 255-274.

Sotomayor, L.R., Cadenillas, A., 2009. Explicit solutions of consumption-investment problems in financial markets with regime-switching. Mathematical finance 19 (2), 251-279.

Wu, H., 2013. Mean-variance portfolio Selection with a stochastic cash flow in a Markov-switching Jump-Diffusion Market. J. Optim. Theory Appl. 158, 918-934.

Yin, G., Zhou, X.Y., 2004. Markowitz mean-variance portfolio selection with regime switching: from discrete-time models to their continuous-time limits. IEEE Transactions Automat. Control 39 (3), 349-360.

Yiu, K.F.C., Liu, J., Siu, T.K., Ching, W.K., 2010. Optimal portfolios with regime switching and value-at-risk constraint. Automatica 46, 979-989.